# Rethinking Thorne–Żytkow Object Formation: The Fate of X-Ray Binary LMC X-4 and Implications for Ultra-long Gamma-Ray Bursts

Tenley Hutchinson-Smith[1,2], Rosa Wallace Everson[1,2,9], Angela A. Twum[1], Aldo Batta[2,3,4], Ricardo Yarza[1,5,10,11], Jamie A. P. Law-Smith[6,7], Alejandro Vigna-Gómez[8], and Enrico Ramirez-Ruiz[1,2]
[1] Department of Astronomy and Astrophysics, University of California, Santa Cruz, CA 95064, USA; tenley@ucsc.edu
[2] Niels Bohr Institute, University of Copenhagen, Blegdamsvej 17, 2100 Copenhagen, Denmark
[3] Instituto Nacional de Astrofísica, Óptica y Electrónica. Tonantzintla, Puebla 72840, Mexico
[4] Consejo Nacional de Humanidades Ciencias y Tecnología, Av. Insurgentes Sur 1582, 03940, Mexico City, Mexico
[5] Texas Advanced Computing Center, University of Texas, Austin, TX 78759, USA
[6] Department of Astronomy and Astrophysics, University of Chicago, Chicago, IL 60637, USA
[7] Kavli Institute for Cosmological Physics, University of Chicago, Chicago, IL 60637, USA
[8] Max-Planck-Institut für Astrophysik, Karl-Schwarzschild-Str. 1, 85748 Garching, Germany
Received 2023 November 15; revised 2024 September 4; accepted 2024 October 17; published 2024 December 12

## Abstract

We present a start-to-end simulation aimed at studying the long-term fate of high-mass X-ray binaries and whether a Thorne–Żytkow object (TŻO) might ultimately be assembled. We analyze results from a 3D hydrodynamical simulation that models the eventual fate of LMC X-4, a compact high-mass X-ray binary system, after the primary fills its Roche lobe and engulfs the neutron star companion. We discuss the outcome of this engulfment within the standard paradigm of TŻO formation. The post-merger angular momentum content of the stellar core is a key ingredient, as even a small amount of rotation can break spherical symmetry and produce a centrifugally supported accretion disk. Our findings suggest the inspiraling neutron star, upon merging with the core, can accrete efficiently via a disk at high rates ($\approx 10^{-2} M_\odot\,\mathrm{s}^{-1}$), subsequently collapsing into a black hole and triggering a bright transient with a luminosity and duration typical of an ultra-long gamma-ray burst. We propose that the canonical framework for TŻO formation via common envelope needs to be revised, as the significant post-merger accretion feedback will unavoidably unbind the vast majority of the surrounding envelope.

*Unified Astronomy Thesaurus concepts:* Transient sources (1851); Close binary stars (254); Interacting binary stars (801); Stellar mergers (2157); Hydrodynamical simulations (767); Late stellar evolution (911); Stellar interiors (1606); X-ray binary stars (1811); High mass x-ray binary stars (733); Gamma-ray bursts (629)

## 1. Introduction

High-mass X-ray binaries (HMXBs) are X-ray emitting sources commonly featuring an early-type massive star and an accreting neutron star (NS; see, e.g., Q. Z. Liu et al. 2006; M. Falanga et al. 2015). A critical juncture in the life of an HMXB is the period just after the primary fills its Roche lobe and engulfs the neutron star companion (B. Paczynski 1976; E. P. J. van den Heuvel 1976). Mass transfer is unstable, and a catastrophic merger with the stellar core cannot be avoided (see, e.g., K. Belczynski et al. 2002; Y. Qin et al. 2019; M. Gallegos-Garcia et al. 2022; C. Liotine et al. 2023). The further evolution of the merger product has been speculated to potentially give rise to a Thorne–Żytkow object (TŻO; K. S. Thorne & A. N. Żytkow 1975, 1977).

A TŻO is a conjectured type of star containing a neutron star at its core, claimed to be formed by the engulfment of a neutron star by an evolving stellar companion (K. S. Thorne & A. N. Żytkow 1975, 1977). TŻOs are speculated to be powered by thermonuclear reactions near the base of the central compact object or by gravitational energy released by mass accretion (G. T. Biehle 1991; R. C. Cannon et al. 1992; R. C. Cannon 1993; G. T. Biehle 1994). In these objects, accretion or thermonuclear heating (and the accompanying radiation) is usually thought to be limited by the self-regulatory balance between gravity and radiation pressure (K. S. Thorne & A. N. Żytkow 1975, 1977; R. C. Cannon et al. 1992). G. T. Biehle (1991) and R. C. Cannon (1993) distinctly showed how in TŻOs with massive stellar envelopes, most of the energy is in fact not provided by accretion onto the NS but by thermonuclear reactions. The fact that TŻOs with massive envelopes generate their luminosities by exotic *rp*-process opens the possibility for stellar abundance measurements to be used to differentiate TŻOs from ordinary evolved stars.

Recently, a chemical anomaly in the red supergiant star HV 2112 in the Small Magellanic Cloud was speculated to be produced by an embedded accreting compact object (E. M. Levesque et al. 2014). The claim that HV 2112 might be a TŻO candidate arises from the anomalously high ratios of calcium, lithium, ruthenium, and molybdenum, which have been predicted to exist inside the exotic accretion-powered stellar interiors of TŻOs. While this claim has incurred numerous rebuttals (C. A. Tout et al. 2014; T. J. Maccarone & S. E. de Mink 2016; E. R. Beasor et al. 2018; A. J. G. O'Grady et al. 2020), it has reinvigorated the construction of numerical models of TŻOs (e.g., R. Farmer et al. 2023), particularly in the context of common envelope (CE) evolution.

While most TŻO calculations suppose a nonrotating, spherically symmetric structure (K. S. Thorne & A. N. Żytkow 1977;

---

[9] NSF Graduate Research Fellow.
[10] NASA FINESST Fellow.
[11] Frontera Computational Science Fellow.

G. T. Biehle 1991; R. C. Cannon et al. 1992; R. Farmer et al. 2023), formation via CE will invariably lead to the inspiraling NS accreting material at rates that are in excess of the Eddington photon limit (R. A. Chevalier 1993, 1996; M. MacLeod & E. Ramirez-Ruiz 2015a). The accretion rate is expected to dramatically increase as the NS sinks dynamically into the center of the star. R. A. Chevalier (1993, 1996) argued that TŻOs may not be able to form during CE since accretion onto the NS might occur in the neutrino-dominated regime. Yet, we note that no fully realistic global simulation of the inspiral phase had been performed.

At the final stage of the stellar merger, the NS ends up being surrounded by dense core material with significant rotational support, whose properties are reminiscent of gamma-ray burst (GRB) progenitors (C. L. Fryer et al. 1999; R. Popham et al. 1999). Understanding the angular momentum content of the surrounding gas is thus critical when thinking about the post-formation evolution of TŻOs. This is because the angular momentum content of the post-merger material determines if a disk will form: if there is no rotation, only Bondi-like accretion will ensue (R. C. Cannon et al. 1992), which, in turn, has important implications for how accretion will advance (W. H. Lee & E. Ramirez-Ruiz 2006; A. Murguia-Berthier et al. 2020; G. Halevi et al. 2023).

Motivated by this, we present here and in a companion paper by R. W. Everson et al. (2024) a study of the long-term fate of HMXBs and whether a near steady-state TŻO might ultimately be produced. In this work, we use the binary system LMC X-4 in the neighboring Large Magellanic Cloud (LMC) satellite galaxy to guide our understanding of such outcomes: LMC X-4 is a highly compact, two-star system consisting of a pulsar and a massive stellar companion and currently has one of the most accurate determinations of the orbital parameters of HMXBs available to date (M. Falanga et al. 2015). LMC X-4 thus appears to be an ideal progenitor candidate for a merged stellar remnant with an embedded neutron star in its core, offering clear constraints for the formation of TŻOs. In a companion paper, R. W. Everson et al. (2024) expands this analysis by exploring the potential formation pathways of TŻOs across a broad parameter space of field binaries comprising an evolved star and stellar-mass compact objects, as well as the expected outcomes for such systems.

This paper is structured as follows. In Section 2, we discuss the role of HMXBs in the context of TŻO candidates arising from stellar mergers involving neutron stars and show using stellar evolution calculations that, as expected, the LMC X-4 binary system will merge via common envelope. Added to that, in Section 2.1 we talk about the fate of the accreting LMC X-4 binary system, which results in the previously accreting neutron star merging with a stellar companion holding to one of the most gravitationally bound envelopes known within the HMXB galactic population. Our stellar evolution models after the primary fills its Roche Lobe are used as initial conditions for 3D hydrodynamical calculations, which are described in Section 3. Our results are then discussed in Section 4, with a summary of this work in Section 5. For completeness, in Appendix B we make use of the results of our global hydrodynamical simulations to show that during the NS inspiral phase, neutrino losses become the dominant energy-loss mechanism, as originally envisioned by R. A. Chevalier (1993, 1996), and the known hydrostatic solutions are not applicable in this regime (K. S. Thorne & A. N. Żytkow 1975; R. C. Cannon 1993).

## 2. The Fate of HMXBs and the Evolution of LMC X-4

### 2.1. Stellar Mergers Involving Neutron Stars and the Role of HMXBs as TŻO Progenitors

The extensively discussed formation scenario for classical TŻOs is that they are the descendants of HMXBs (R. E. Taam et al. 1978; R. C. Cannon 1993; P. Podsiadlowski et al. 1995). When in these binary systems, the massive primary fills its Roche lobe, unstable mass transfer ensues, and the system will likely evolve into a CE phase, where the NS is completely engulfed by the envelope of the evolving companion (R. W. Everson et al. 2024). Due to gas drag (M. MacLeod & E. Ramirez-Ruiz 2015a), the NS will then naturally spiral toward the primary's core. If the orbital energy released in the sinking is enough to eject the envelope, the end product will likely give rise to a merging compact binary (T. Fragos et al. 2019). If, on the other hand, the orbital energy released is not sufficient to eject the envelope, the NS will sink into the core of the primary star and could become a TŻO (R. E. Taam et al. 1978).

The formation of classical TŻOs, as argued by R. W. Everson et al. (2024) in a companion paper, is likely to be aided by the engulfing primary being lighter and early on its evolution. The reasons for this are twofold. First, the stellar envelope is more highly bound and thus less likely to be ejected during CE. Second, the core is less dense, which naturally gives rise to lower mass accretion rates once the NS has settled into the primary's core. Because of this, here we study the long-term fate of HMXBs and scrutinize whether a near steady-state TŻO might ultimately be produced.

Figure 1 shows the population demographics of binary systems at the onset of CE, constructed using the library generated using the rapid binary population synthesis code COMPAS version 02.27.05 (J. Riley et al. 2022) at $Z = 0.142$ and with CE efficiency parameter set to unity. As expected, HMXBs are included in the population of CE events with NS companions that are expected to merge (i.e., unable to eject the entire envelope).

Only in a handful of sources among the more than 100 HMXBs is the inclination of the binary along our line of sight high enough for the NS to be periodically obscured by the orbiting companion. This naturally gives rise to X-ray eclipses, which can be used to constrain a wide range of properties of the binary system. Figure 1 shows the properties of the stellar companions in the eclipsing HMXB systems studied by M. Falanga et al. (2015). LMC X-4 is particularly noteworthy. It has a relatively moderate mass and a highly bound envelope as well as currently having one of the most accurate determined orbital properties. As such, LMC X-4 appears to be an ideal TŻO progenitor candidate.

### 2.2. Forward Modeling the Pre-merger Evolution of LMC X-4

In order to explore the potential for TŻO formation during the expected merger of the LMC X-4 system, we develop initial conditions based on the best current mass, size, and kinematic constraints available (M. Falanga et al. 2015), forward-evolve them with MESA to common envelope onset, then map the resulting future system into 3D hydrodynamics and evolve it through CE and merger.

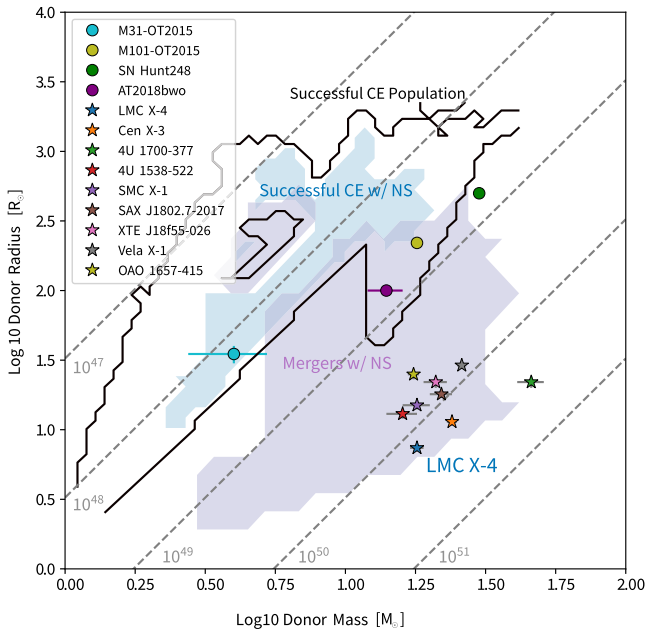

**Figure 1.** HMXBs in the context of luminous red novae (LRNe) and binary populations synthesis. The masses and radii of eclipsing HMXB systems (stars) studied by M. Falanga et al. (2015) and pre-CE transient observations for a sample LRNe progenitors (T. Matsumoto & B. D. Metzger 2022) are depicted (circles). Also shown are the progenitor population demographics of CE events generated with the rapid binary population synthesis code COMPAS version 02.27.05 (J. Riley et al. 2022). The COMPAS populations of primary stars include all systems involved in successful CE events, for which most companions are stars and white dwarfs (black outline), as well as CE events with NS accretors that either successfully ejected the envelope (blue region) or merge (purple region). The gravitational binding energy (in units of ergs) of a polytrope of index $n = 3$ with mass $M$ and radius $R$ (gray dashed lines) is shown for guidance.

### 2.3. Properties of LMC X-4

Over a decade of monitoring with multiple instruments (Q. Z. Liu et al. 2006; M. Falanga et al. 2015) has provided strong constraints on the properties of LMC X-4. According to M. Falanga et al. (2015), this HMXB system comprises a $1.57 \pm 0.11 M_\odot$ NS that is accreting from its $18 \pm 1 M_\odot$ stellar companion. The companion, hereafter referred to as the primary, somewhat underfills its Roche lobe with a stellar radius of $7.4 \pm 0.4 R_\odot$. The system has a separation of $14.2 \pm 0.2 R_\odot$ and an observed orbital period of 1.408 days. M. Falanga et al. (2015) estimated the orbital period change $\dot{P}/P$ as $-1.0 \times 10^{-6}$ yr$^{-1}$, suggesting a gradual tightening of the orbit. The estimated mass-loss rate of the primary due to winds is $2.4 \times 10^{-7} M_\odot$ yr$^{-1}$ (S. Chernov 2020), which alone cannot account for the estimated $\dot{P}/P$ value.

### 2.4. Stellar Models

We begin with modeling the evolution of the primary from the present time to the onset of CE[12] using the MESA Isochrones and Stellar Tracks (MIST) package (J. Choi et al. 2016; A. Dotter 2016) with MESA v7503 (B. Paxton et al. 2011, 2013, 2015). The MIST framework is chosen due to its calibration to observations, including in the LMC. We evolve the primary as a single star using the mean LMC metallicity of [Fe/H] $= -0.42$ (S. Choudhury et al. 2021). We choose our initial mass range $[16, 20] M_\odot$ to be consistent with the $2\sigma$ mass estimates from M. Falanga et al. (2015). This allows us to create a grid of potential models that will be narrowed down based on the properties of the primary star.

---

[12] The in-lists utilized are available on Zenodo at doi:10.5281/zenodo.13634535.

Because the evolution of the orbital period is not totally independent of the evolution of the primary, we consider both. Prior work has attempted to explain the $\dot{P}/P$ value of LMC X-4 with conservative mass transfer (S. Safi-Harb & H. Ögelman 1996) and changes in the primary's moment of inertia (A. M. Levine et al. 2000). Upon revisiting these works, in the former case, the mass-transfer rate required is much higher than the observationally derived mass-loss values for the primary, and in the latter case, the formalism cannot match the measured $\dot{P}/P$ when applied to up-to-date stellar models in the appropriate stage of evolution (S. Chernov 2020). For this system, wind mass loss serves only to make orbital tightening more difficult.

Recent work by S. Chernov (2020), however, demonstrates that tides alone are sufficient to explain LMC X-4's orbital tightening. Though the rate of orbital tightening does include effects from winds, mass transfer, and the primary's change in moment of inertia, it's not possible to completely distinguish these effects with those of the degree of corotation on dynamical tides. This leaves LMC X-4's separation as a function of time $a(t)$ as our last additional constraint on the primary, as the Roche lobe acts as a size constraint and trigger of mass loss for our models. The models with initial masses of $18 M_\odot$ and $19 M_\odot$, with mass loss as prescribed by S. Chernov (2020), are consistent with the properties given by M. Falanga et al. (2015) within $1\sigma$, and as the 3D hydrodynamical results for both are qualitatively similar, we present those from the $18 M_\odot$ model below.

Using the size of the primary's Roche lobe, we allow the models to lose mass as they approach the end of the main sequence via winds and a radius-limited prescription for Roche lobe overflow. When the model reaches the observed mass and radius constraints (black symbol in Figure 2), we mark that model as "today," then allow the star to continue its evolution while evolving the separation $a$ using the observed $\dot{P}/P$. We then map the resulting "future" model, in which the donor has overfilled its Roche lobe and its radius equals the forward-evolved separation $a$ (turquoise symbol in Figure 2), into 3D for the simulation of CE and merger. The reader is referred to Appendix A for additional details on the interior properties of the stellar companion MESA model.

## 3. Studying the Fate of X-Ray Binary LMC X-4 with Hydrodynamical Simulations

### 3.1. FLASH Setup

We map the properties of the MESA profile (density, pressure, temperature, and composition) onto a 3D grid using FLASH version 4.32 (B. Fryxell et al. 2000). Our numerical scheme is adapted from S. Wu et al. (2020), which was originally developed by J. Guillochon et al. (2009) and J. Guillochon & E. Ramirez-Ruiz (2013), and subsequently upgraded by J. Law-Smith et al. (2019), J. A. P. Law-Smith et al. (2020a, 2020b). The upgraded version uses an extended Helmholtz equation of state (F. X. Timmes & F. D. Swesty 2000) to track individual elements as described in J. Law-Smith et al. (2019) and S. Wu et al. (2020). The reader is referred to these works for further numerical details. A brief summary including key aspects and alterations to the setup is given below.

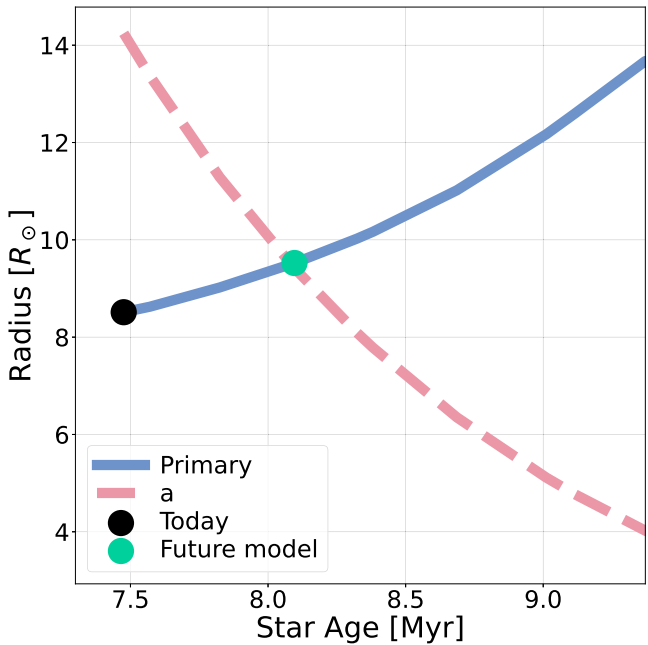

**Figure 2.** The predicted radial evolution of the primary star as calculated by MESA using the mean LMC metallicity of [Fe/H] = −0.42. The initial mass of the model is $18M_\odot$, with mass loss $2.4 \times 10^{-7} M_\odot$ yr$^{-1}$ (S. Chernov 2020). The selected model (black symbol) is consistent with the properties given by M. Falanga et al. (2015) within $1\sigma$. The model used for the hydrodynamical simulation performed with FLASH is marked with a turquoise symbol in the panel. At this stage, the donor has overfilled its Roche lobe, and its radial scale matches the forward-evolved separation $a$ (dashed line).

To set up the initial model, we first relax the MESA profile in FLASH for a few dynamical timescales. The computational domain is cubic with volume $(60R_\odot)^3$ and comprises an $8^3$ block grid with a minimum cell size of $0.02R_\odot$. During the relaxation process, a point mass, representing the NS with $M_{\rm ns} = 1.57M_\odot$, is introduced at $9R_\odot$. The primary star has a radial scale $\approx 9.4R_\odot$ when it overfills its Roche lobe (Figure 2). The NS is initially at rest but its velocity is slowly augmented as the relaxation process ensues until reaching a circular Keplerian velocity. At the time the relaxation finishes, the model of the primary star is in hydrostatic equilibrium, and the inspiral trajectory of the NS is calculated self-consistently (S. Wu et al. 2020). The properties of the merger remnant are found to be rather indifferent to the exact initial conditions of the NS's velocity, as long as it is close to Keplerian. We stop the simulation after the NS sinks into the center of the primary's core ($t = 13.8$ hr).

### 3.2. Dynamical Inspiral

Now we turn our attention to LMC X-4 at the onset of CE and beyond. As the engulfment proceeds (from left to right in the top three panels of Figure 3), the NS plunges dynamically into the stellar core of the mildly evolved primary star. This steep plunge-in is driven by strong drag forces, which are significantly higher than those predicted by Bondi-Hoyle-Lyttleton accretion theory (F. Hoyle & R. A. Lyttleton 1939) when the stellar density gradients are included, as shown by M. MacLeod & E. Ramirez-Ruiz (2015a, 2015b), M. MacLeod et al. (2017a), and R. W. Everson et al. (2020). We compare the numerical trajectories with the modified drag predictions presented in M. MacLeod & E. Ramirez-Ruiz (2015a) and find rough agreement with the dynamical plunge seen in Figure 3.

As can be seen from Figure 3, the NS disturbs a nonnegligible fraction of the outer stellar envelope during inspiral. It then crosses the core without disrupting it and settles inside the stellar core after about 13.8 hr. As a result of the merger, the NS injects energy and angular momentum (via spiral waves) into the primary and unbinds a small fraction of the envelope material. The bottom panels in Figure 3 show specific energy, $\varepsilon = \varepsilon_{\rm kin} + \varepsilon_{\rm grav}$ (the sum of specific kinetic and potential energy while the internal energy is not included). The orange region corresponds to unbound material ($\varepsilon > 0$) while the green region corresponds to bound material ($\varepsilon < 0$). At early times, $\varepsilon < 0$ for most cells in the box. At the final time depicted in Figure 3, a nonnegligible fraction (see Section 3.3 for details) of the material in the box (except for the surviving core) has $\varepsilon > 0$. The reader is referred to Figure 14 in Appendix A for a meridional view of the dynamical plunge seen in Figure 3. In what follows, we engage in a more comprehensive evaluation of the ejection of material during the merger.

### 3.3. Mass Ejection during Merger

Figure 4 shows, as a function of the mass coordinate, the binding energy of the envelope and the change in the orbital energy from the start of the inspiral, using the properties of the stellar model that we have selected as the initial condition for the hydrodynamical simulation. This shows the necessary conditions for effective envelope ejection, which are expected to be satisfied when the change in orbital energy of the NS, lost through drag during inspiral, is similar to the envelope's binding energy at a particular mass coordinate, known as the $\alpha = (E_{\rm grav}/\Delta E_{\rm orb}) = 1$ condition. The calculations presented in Figure 4 predict that the NS would unbind $\approx 0.4 M_\odot$ of envelope material. Note that although this calculation is only approximate, it allows us to better understand the results of a detailed simulation able to capture the energy sharing during the inspiral process.

Since the orbital energy deposited by the NS through inspiral is less than the gravitational binding energy of the stellar envelope, a merger is an inevitable outcome for LMC X-4. As the NS inspirals, the structure of the envelope internal to the position of the NS will remain fairly unaltered, and the orbit of the NS will shrink until it merges with the stellar core.

We first note that the change in orbital energy continues to slightly increase although it remains well below the envelope's binding energy for the region between the core and the $\alpha = 1$ crossing point. As seen in Figure 3, the envelope is shocked and swept preferentially outward as the NS moves through the envelope of the primary. Basically, the NS acts as a local diffusive energy source term, giving surrounding material roughly outward radial velocities. By the time the engulfed NS reaches the outer edge of the stellar core, we find $\approx 1.4M_\odot$ of mass to be unbound in our simulation (Figure 5) by summing the mass of the cells with $\varepsilon > 0$ at the $t = 4.5$ hr snapshot of our hydrodynamic simulation. This time is chosen as the moment when the NS reaches the core, and no envelope material has yet been ejected from the computational domain. The velocity vectors of each grid cell that meets this criterion at $t = 4.5$ hr are nearly all pointed radially outward. The outward radial velocity distribution of the envelope material expelled during the merger episode is shown in Figure 5. The distribution shown in Figure 5 is compared to the local escape velocity at the $\alpha = 1$ critical radius, $v_{\alpha = 1} \approx 1140$ km s$^{-1}$, which is computed based on the enclosed mass of the initial model (Figure 4). The deposited energy is not shared efficiently throughout the envelope (M. MacLeod et al. 2022) as the orbit of the NS shrinks and, as a result, shocks accelerate a nonnegligible fraction of the envelope mass to well above the escape velocity.

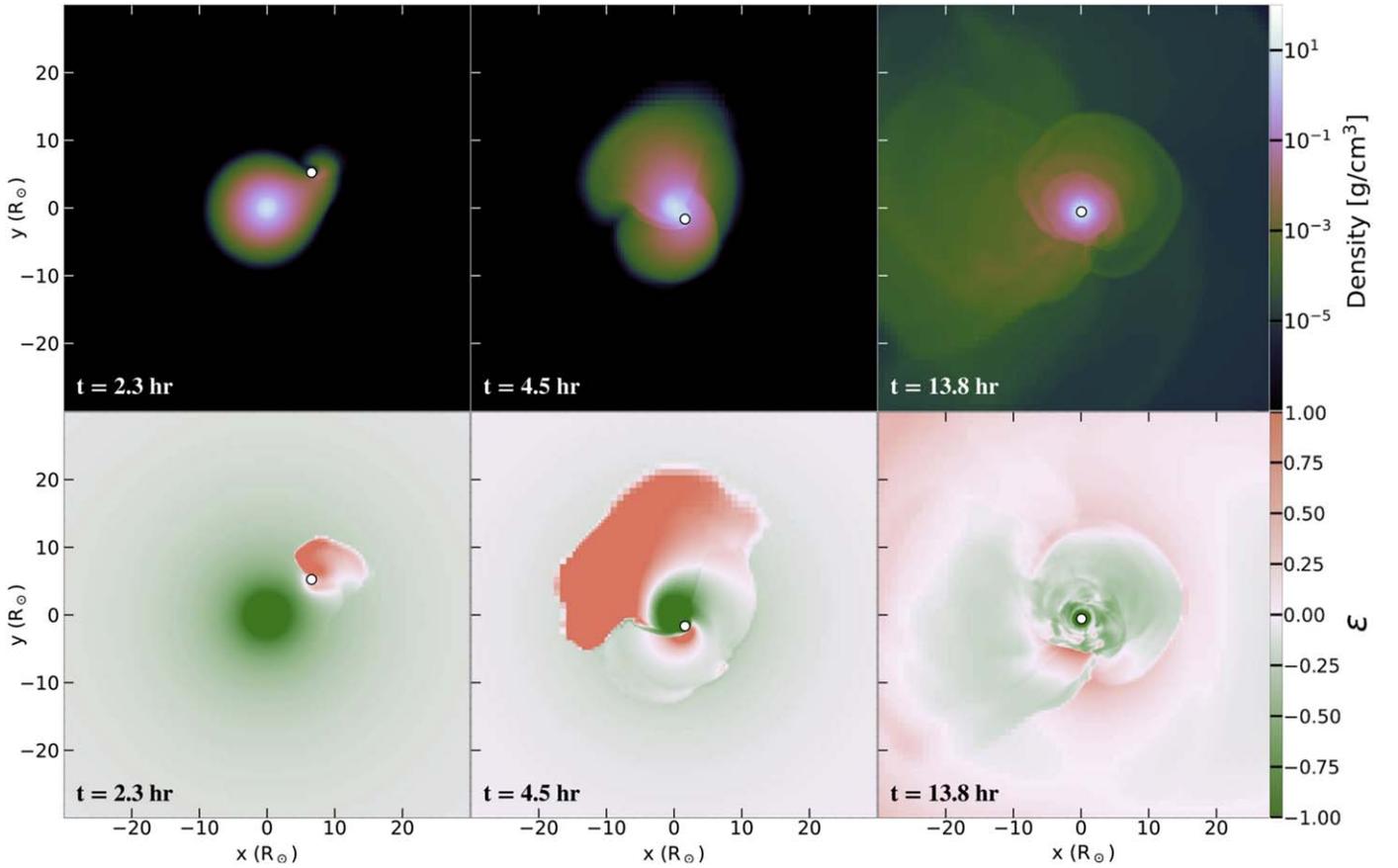

**Figure 3.** Two-dimensional slices across the orbital plane of FLASH simulated inspiral and merger of LMC X-4 at three stages: early in the evolution (2.3 hr), at an intermediate time when the NS enters the stellar core (4.5 hr), and at a late time (13.8 hr) after the NS has settled in the center of the stellar core. Top panels: logarithm of gas density. Bottom panels: sum of specific kinetic and potential energy, $\varepsilon = \varepsilon_{\rm kin} + \varepsilon_{\rm grav}$, in units of $10^{16}$ cm$^2$ s$^{-2}$. The orange region corresponds to unbound material ($\varepsilon > 0$), and the green region corresponds to bound material ($\varepsilon < 0$).

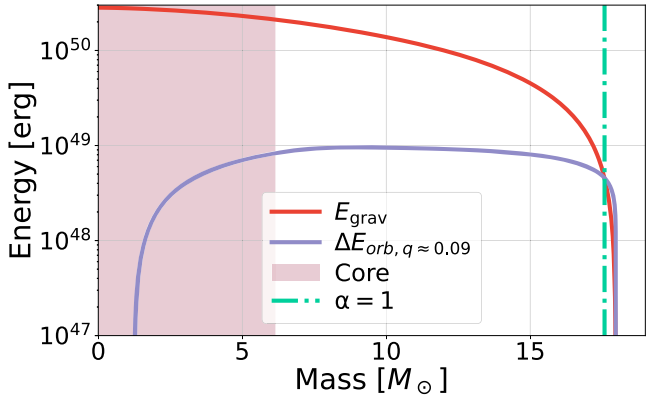

**Figure 4.** Relevant quantities required to estimate envelope unbinding during the merger, which are shown in mass coordinates. The binding energy of the primary's stellar material exterior to a given mass coordinate ($E_{\rm grav}$, red line) and the change in orbital energy expected to be dissipated during the inspiral ($\Delta E_{\rm orb}$, purple line) are plotted against the mass coordinate for the model used as the initial condition for the hydrodynamical simulation, with mass $18 M_\odot$ and neutron star mass ratio $q \approx 0.09$. The mass coordinate at which the $\alpha = (E_{\rm grav}/\Delta E_{\rm orb}) = 1$ condition is satisfied is shown as a vertical dashed line.

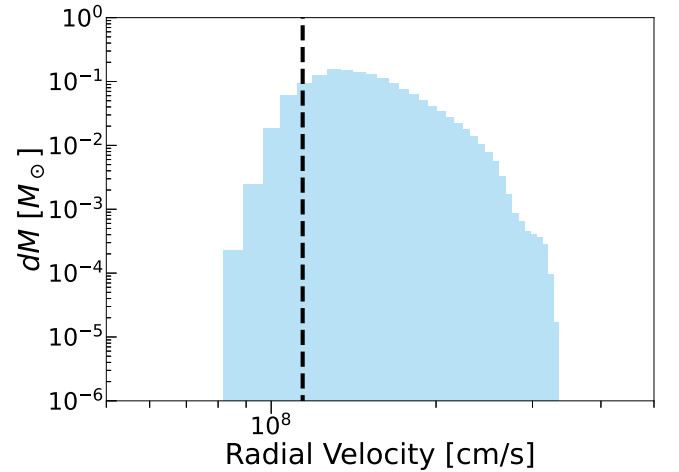

**Figure 5.** The outward radial velocity histogram of the envelope debris expelled during the merger (with a total mass $\approx 1.4 M_\odot$). This is done by analyzing the velocity vectors and masses of each grid cell in the envelope with $\varepsilon > 0$ at $t = 4.5$ hr. Also plotted is the local escape velocity at the $\alpha = 1$ critical radius, which is computed based on the enclosed mass of the initial model (Figure 4). This velocity is $v_{\alpha=1} \approx 1140$ km s$^{-1}$ and is shown as a vertical dashed line.

We conclude that the $\alpha = 1$ condition provides a lower limit to the amount of energy injected into the stellar envelope. Some of the excess energy resides in the thermal energy content stored in the primary star (T. Fragos et al. 2019), which, as the envelope expands, does work. In addition, as the NS continues to sink below the $\alpha = 1$ critical radius, a larger fraction of orbital energy can be injected into the envelope (S. Wu et al. 2020). As such, we must consider the internal energy (e.g., T. Fragos et al. 2019), the remaining orbital energy as the NS continues to sink (e.g., J. A. P. Law-Smith et al. 2020b; S. Wu et al. 2020), and the anisotropic sharing of this energy in the envelope (e.g., M. MacLeod et al. 2018) in order to better estimate the properties of the outflow. The outcomes of this study have direct consequences for the efficiency of envelope ejection, which is directly relevant to astronomical transients powered by recombination (e.g., M. MacLeod et al. 2017b; S. L. Schrøder et al. 2020).

### 3.4. On the Accretion-fed Growth of the Neutron Star during the Merger Phase

The accretion-fed growth of NSs during CE events has been discussed extensively in the context of successful CE ejection. In these cases, possible outcomes include close binaries comprising an He-star and an NS, which might leave behind a double NS or NSBH binary (K. A. Postnov & L. R. Yungelson 2014). Accretion onto the NS during this phase was thought to be effectual, as neutrinos provide a cooling mechanism (J. C. Houck & R. A. Chevalier 1991; R. A. Chevalier 1993; C. L. Fryer et al. 1996), suggesting that an NS might undergo accretion-induced collapse to a BH in some, but not all, cases (R. A. Chevalier 1993; G. E. Brown 1995; H. A. Bethe & G. E. Brown 1998).

The idea that an NS might be able to grow via accretion during CE is hard to reconcile with the observed distribution of NS masses (e.g, F. Özel et al. 2012) and, most notably, with those inferred in double NS binaries. These systems show a narrow range of inferred masses centered at $1.33 M_\odot$ with dispersion of $0.05 M_\odot$ (F. Özel et al. 2012). M. MacLeod & E. Ramirez-Ruiz (2015a) demonstrated that the presence of a density gradient in stellar envelopes significantly restricts accretion by imposing a net angular momentum to the flow around the NS, and that embedded NSs should accrete only modest amounts of envelope material. This study presented a clear hydrodynamical solution that successfully explains the narrow mass distribution seen in double NS binaries. This was later confirmed by J. A. P. Law-Smith et al. (2020b) with the use of global hydrodynamical simulations. These studies have been, however, restricted to the survival of NSs in successful CE events. Further work is needed to probe the efficiency of accretion in NSs that are unable to eject the envelope and, as a result, merge with the dense core.

We now extend these calculations to consider the accretion history of the NS in LMC X-4. In Figure 6 we plot the mass accretion rate onto the inspiraling NS before reaching the core (dashed vertical line). We find that the NS is able to gain about $0.14 M_\odot$ and the inferred accretion rate across the point mass remains above the neutrino-cooling rate of $\dot{M}_\nu \approx 10^4 \dot{M}_{\rm Edd} \approx 10^{-4} M_\odot\ {\rm yr}^{-1}$ (J. C. Houck & R. A. Chevalier 1991) at which accretion growth can proceed effectively. The results presented in Figure 6 are in agreement with the mass accretion rates derived using the formalism of M. MacLeod & E. Ramirez-Ruiz (2015a) for this specific progenitor. A limitation of most numerical models is that the accretion rate depends sensitively on the size of the central absorbing sink (e.g., A. Murguia-Berthier et al. 2017a; S. De et al. 2020), which implies that the results presented in Figure 6 are upper limits. This mass gain, nonetheless, represents a small fractional amount and will not drastically

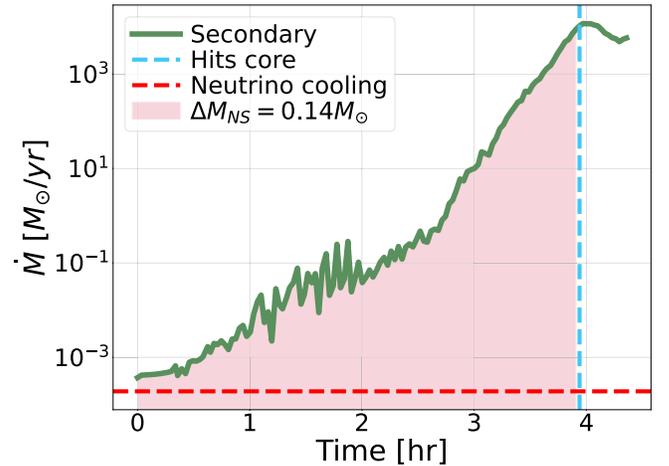

**Figure 6.** Mass accretion rate onto the inspiraling NS as a function of time during the simulation shown in Figure 3. In compact primaries, the NS is able to accrete above the limit at which neutrinos are able to provide the main cooling agent (red dashed line). In LMC X-4, the NS is able to gain about $0.14 M_\odot$, but remains well below the maximum NS mass until merging with the core (blue dashed line).

affect the structure of the NS, which will remain well under the maximum mass limit.

### 3.5. Embedded Phase and Merger with the Stellar Core

Here, we take a closer look at the final plunge of the NS and its impact on core structure and final merger. Figure 7 provides a zoomed-in view of the stellar core near the end of the inspiral. The relevant quantity fixing the strength of the interaction is $q_c^{1/3} = (M_{\rm NS}/M_{\rm core})^{1/3}$ (R. W. Everson et al. 2024) and, as such, we expect the NS to only mildly impact the stellar core. This is because $q_c < 1$, making the extent of the core larger than the tidal radius with respect to the NS (R. W. Everson et al. 2024). Shown in Figure 7 are illustrations of the gas density of the primary's core at two different times, where the top panels show slices in the orbital plane (y–x), and the bottom panels display slices in the vertical plane (y–z). The two inset panels in Figure 7 show the trajectory of the NS in the orbital plane from the moment it reaches the core until the end of the simulation.

The core of the primary spins up due to shocks generated during inspiral, absorbing orbital angular momentum deposited by the NS. If the spin of the core material is lower than that required to form a disk, we can assume quasi-spherical accretion, which is necessary to power a classical TŻO (R. W. Everson et al. 2024). If, on the other hand, the spin of the core is higher than that required to form a disk, significant post-merger accretion feedback is expected to unbind the vast majority of the remaining envelope (S. S. Bavera et al. 2020; A. Murguia-Berthier et al. 2020). The post-merger angular momentum content of the stellar core is thus essential to defining the type of merger product that will result. Motivated by this, our aim in Section 4 is to discuss the post-merger flow pattern around the NS, involving accretion, rotation, and directional outflow.

## 4. Implications for Thorne–Żytkow Object Formation and Ultra-long Gamma-Ray Bursts

We now explore the final outcome of LMC X-4 as informed by our simulations. As a result of the merger, the NS injects energy, mass, and angular momentum into the core of the primary star (Figure 3) and unbinds a small amount of envelope material in the process (Figure 5). Shocks in the course of the inspiral spin up the core before the NS ultimately settles into the core's interior (Figure 7). LMC X-4 thus naturally evolves into a collapsar-like progenitor system (C. L. Fryer & S. E. Woosley 1998)—that is, a hyper-accreting NS surrounded by a rotating stellar core.

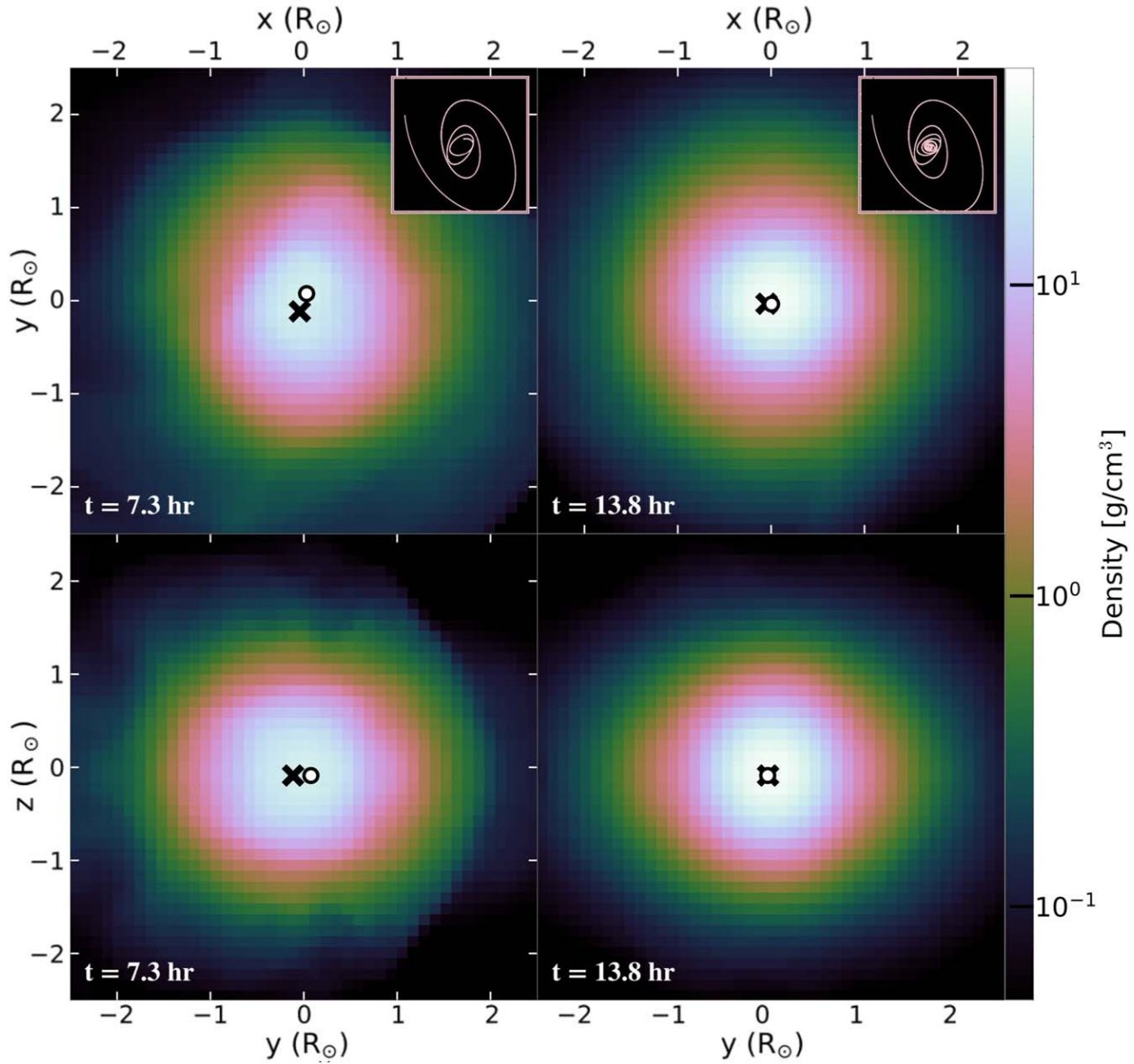

**Figure 7.** Two-dimensional slices of the logarithm of gas density at two different times (7.3 hr and 13.8 hr) for material near the core of the primary. The top panels show slices in the orbital plane (y–x) while the bottom panels display slices in the meridional plane (y–z). The position of the core's center of mass is denoted by an X mark, while the position of the NS is denoted by a white circle. The evolution of the trajectory of the NS in the orbital plane from the time the NS reaches the core is shown in the respective inset panels with a characteristic height and width of $(4R_\odot)^2$.

### 4.1. A Rapidly Rotating Stellar Core Accreting onto a Central Neutron Star

The first task in attempting to construct a general scheme that is able to successfully describe the fates of X-ray binaries such as LMC X-4 is to decide which parameters exert a controlling influence upon their final properties. The mass of the central object is a key parameter, as it dictates a characteristic luminosity scale for the ensuing accretion activity (e.g., C. L. Fryer & S. E. Woosley 1998) as well as determining the post-merger properties of the stellar core (R. W. Everson et al. 2024). The total angular momentum of the core may be, however, more important in this regard as it determines whether or not an accretion disk will be formed immediately upon merger.

Figure 8 displays the specific angular momentum content in the stellar core, calculated by taking the spherical average about the center of mass of the FLASH simulation shown in Figure 7. The specific angular momentum is shown in units of $j_{\rm isco}$ (J. M. Bardeen et al. 1972), the minimum specific angular momentum required for disk formation, and is plotted against the enclosed mass, which includes the NS. As expected, the specific angular momentum is above the critical limits $j/j_{\rm isco} = 1$ at all mass coordinates. We expect a BH to be formed when the NS reaches a mass of about $3M_\odot$ (pink shaded region in Figure 7). After the inevitable accretion-induced collapse of the NS, the expected outcome would then be a spinning BH surrounded by an accretion disk, which is reminiscent of central engine models for GRB sources (R. Mochkovitch et al. 1993; S. E. Woosley 1993; C. L. Fryer & S. E. Woosley 1998; A. I. MacFadyen & S. E. Woosley 1999; M. J. Rees 1999; M. A. Aloy et al. 2000; C. L. Fryer & A. Heger 2000; A. I. MacFadyen et al. 2001; W. H. Lee & E. Ramirez-Ruiz 2007; N. Gehrels et al. 2009; O. Gottlieb et al. 2022).

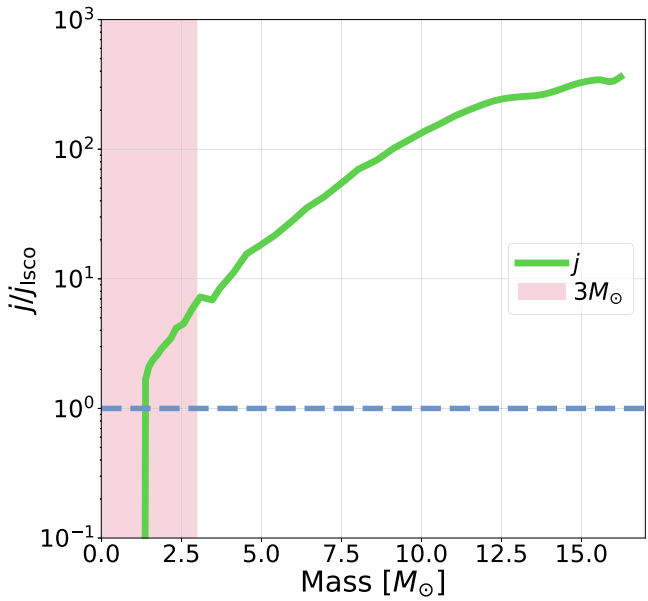

**Figure 8.** The normalized specific angular momentum content $j/j_{\rm isco}$ of the stellar core at the end of the simulation is shown in mass coordinates (green curve). The profile is calculated by taking the spherical average about the center of mass in the FLASH simulation shown in Figure 7. Here, $j_{\rm isco}$ is the specific angular momentum required for disk formation (dashed vertical line). We note that the simulated stellar model satisfies this condition for disk formation at all radii.

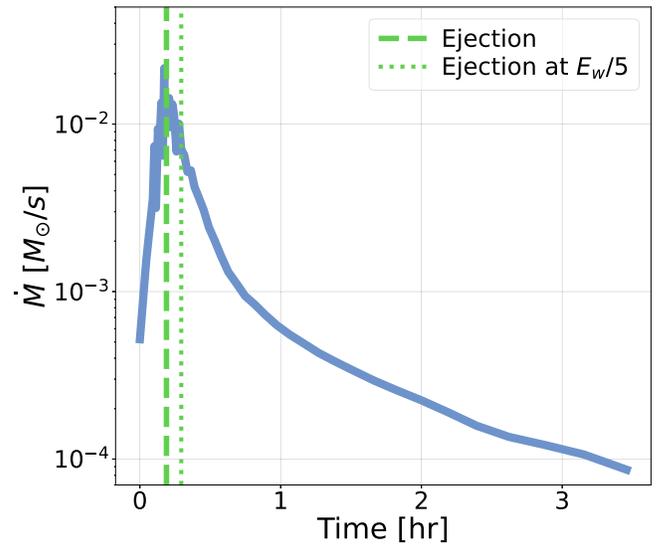

**Figure 9.** The accretion rate onto the newly formed BH as a function of time. The rotating infalling core material is assumed to freefall without any pressure support and without mixing of radial layers. A disk is presumed to form from the collapse of every infalling shell and then subsequently accreted before the next shell collapses, as described in S. S. Bavera et al. (2020). When feedback from the accretion disk is taken into account, we show that it can unbind the infalling material on timescales that are comparable to the peak accretion time. The two ejection timescales shown as vertical lines are introduced in Figure 10.

To calculate the accretion rate history of the accreting BH, we follow the framework described in A. Batta & E. Ramirez-Ruiz (2019) and S. S. Bavera et al. (2020). Following the assumption that a BH is formed from the collapse of the innermost $3M_\odot$ of the stellar core, we presume that the rotating infalling material free falls onto the recently formed BH without any pressure support and without mixing of radial layers. In this formalism, the mass distribution $M(r)$ of the stellar core can be envisioned as a collection of shells with mass $m_{\rm shell}$ and angular frequency $\Omega_{\rm shell}$ that fall one after the other. Assuming that the disk formed from the collapse of a shell ($m_{\rm shell}$ with $j > j_{\rm isco}$) is accreted before the next shell collapses, as the viscous accretion timescale of the inner disk is shorter than the dynamical timescale of the collapsing shells, we can calculate the accretion rate onto the BH, while consistently evolving the BH's mass and spin as material is amassed. The reader is referred to S. S. Bavera et al. (2020) for further details.

The resulting mass accretion rate is displayed in Figure 9. The peak accretion luminosity is, as expected, well below that expected for most collapsar models ($\gtrsim 10^{-1} M_\odot$ s$^{-1}$; C. L. Fryer & S. E. Woosley 1998; A. I. MacFadyen & S. E. Woosley 1999), and the central engine is foreseen to be longer lived that the typical duration of long GRBs ($\approx$20 s; N. Gehrels et al. 2009). This is expected, as the core of the primary star is not highly evolved and thus remains much less compact at the time of collapse.

The accretion activity might be, however, shortened when the feedback energy from the wind's accretion luminosity is taken into consideration. It has been widely shown that general relativistic, magnetohydrodynamic simulations of accretion disks around rotating BHs (e.g., J. C. McKinney et al. 2012) commonly show the development of a magnetized wind outflow, which can introduce energy that can, in this case, be competently shared with the layers of the infalling core. Figure 10 shows how the accumulated energy injection from the wind may contribute to envelope ejection as the stellar material infalls and subsequently accretes. The feedback from accretion is assumed to launch a wind that shares energy $E_{\rm w}$ with the collapsing stellar material with an efficiency of 1% (0.2%). The envelope is expected to be ejected when the integrated energy from the wind is larger than the binding energy of the envelope (see the dashed lines in Figure 10). This provides us with a rough idea of the total energy available to unbind the star. The feedback energy from accretion would then stop the collapse and notably abbreviate the duration of the accretion event.

### 4.2. Ultra-long GRBs and Recombination Transients

In this Section, we present a short summary of what we have learned so far about the fate of LMC X-4. Figure 11 presents a schematic montage of the events in the life of the LMC X-4 system. This encourages us to present a preliminary account of the effects of the ensuing accretion activity onto the central compact object. We also briefly describe the observational prospects for the near future. The NS at the center of the merger remnant quickly accretes a sizable amount of mass and is expected to collapse to a BH. The expected outcome would then be a spinning BH, orbited by a torus of stellar core debris. The binding energy of the orbiting debris and the spin of the BH are the two main energy reservoirs. It has become increasingly conspicuous that most plausible GRB progenitors (e.g., M. J. Rees 1999; R. G. Izzard et al. 2004; W. H. Lee & E. Ramirez-Ruiz 2007; N. Gehrels et al. 2009; A. Murguia-Berthier et al. 2017b) are expected to lead to a BH plus debris torus system with the overall energetics mainly determined by the mass and spin of the BH and the different masses left behind in the orbiting debris. This accretion process is expected to power and launch a relativistic jet, which will pierce through the primary star (e.g., A. I. MacFadyen et al. 2001; E. Ramirez-Ruiz et al. 2002; O. Gottlieb et al. 2022). Energy dissipation within the jet is expected to produce γ-ray emission, which will eventually be detected as a GRB for on-axis observers.

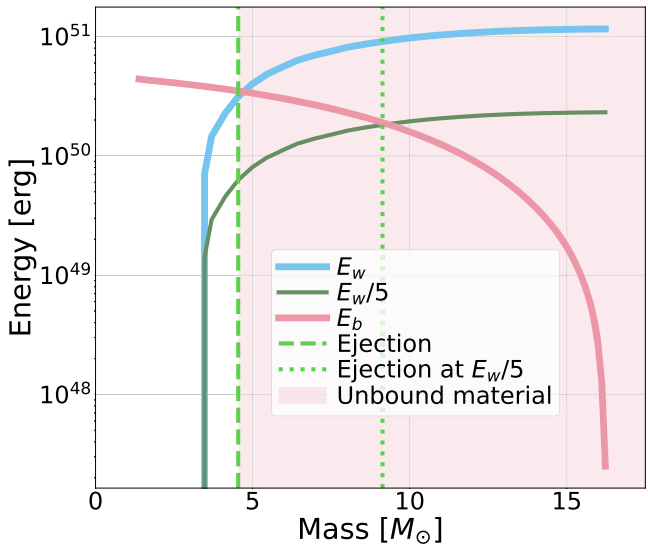

**Figure 10.** Relevant energy scales needed to estimate envelope unbinding during the accretion phase onto the newly formed BH. The binding energy of the rotating collapsing envelope, $E_b$, exterior to a given mass coordinate (pink) and the integrated wind accretion energy, $E_w$, expected to be shared with the infalling envelope (blue and dark green) are plotted against the mass coordinate of the post-merger model. The integrated energy required for ejection is uncertain. Here we assume two feedback efficiencies when calculating the fraction of energy that is shared with the collapsing stellar material. These are 1% (blue) and 0.2% (dark green), respectively. The formation of an accretion disk induces feedback, which in turn imposes a limit to the fraction of the star that can collapse and power an accretion event (dashed and dotted vertical lines).

The expected properties, including luminosities and decay timescales, are consistent with properties of ultra-long GRBs (ULGRBs; e.g., A. J. Levan et al. 2014). These ULGRBs attain peak X-ray luminosities of $\approx 10^{49}$ erg s$^{-1}$, have a duration on the order of a few hours, and have nonthermal spectra evocative of relativistically beamed emission. In Figure 12, we plot the peak timescale and luminosity of the high-energy transient following the merger of LMC X-4. We calculate the beamed luminosity of the jetted transient by assuming it traces the mass supply onto the BH, $L \propto \dot{M} c^2$ (Figure 9). We also assume that 1% of the accretion luminosity goes into high-energy radiation. It is thus enticing to identify ultra-long GRBs with the remarkable merger of a compact object and its stellar companion, with a small fraction of accretion energy going into a jetted outflow. This could be the less extreme example of a collapsar progenitor (S. E. Woosley 1993; A. I. MacFadyen & S. E. Woosley 1999).

A potential impediment for such relativistic outflows is the amount of entrained baryonic mass from the surrounding environment. In GRB models, high neutrino fluxes are capable of ablating baryonic mass from the surface of the disk at a rate $\dot{M}_{\rm wind} \propto L_\nu^{5/3}$ (Y. Z. Qian & S. E. Woosley 1996). Here, $L_\nu$ is the neutrino luminosity, which, in turn, is expected to decrease with the mass accretion rate. Assuming that the poloidal field strength is restricted by the vigor of the convective motions, the magnetic luminosity scales as $L_{\rm wind} \propto B^2 \propto L_\nu^{2/3}$ (W. H. Lee & E. Ramirez-Ruiz 2007). As a result, the limiting bulk Lorentz factor of the outflow, $\Gamma \approx L_{\rm wind}/(\dot{M}_{\rm wind} c^2)$, decreases as $L_\nu^{-1}$ for neutrino-dominated accretion disks (R. Narayan et al. 2001). This reveals that the entrained baryonic mass is expected to be less severe in TŻO than GRB environments.

If we were to venture a general classification scheme for the fate of HMXBs, on the hypothesis that the central engine involves a BH formed in the collapse of post-merger stellar cores (R. W. Everson et al. 2024), we would surely expect the mass of the BH, the rate at which the gas is supplied to the BH (which will depend on the mass and evolutionary state of the primary star), the spin of the BH, and the orientation relative to our line of sight to all be essential parameters. For off-axis observers, we expect the mass ejected from the merger and the ensuing feedback to power an optical or infrared transient as it expands and becomes transparent (N. Soker & R. Tylenda 2006; B. D. Metzger et al. 2012; N. Ivanova et al. 2013; M. MacLeod et al. 2017b). The detection of this post-merger transient would offer direct constraints on the properties of the outflow, such as the total mass and velocity distribution. We expect that the ensuing accretion feedback, which will likely halt accretion, would help unbind most of the stellar envelope and thus produce a much brighter recombination transient than the one expected from the outflow conditions shown in Figure 5.

### 4.3. Previously Discussed Transients Arising from TŻOs

Although some of the features discussed in Section 4.2 were anticipated by previous theoretical discussions, the development and use of hydrodynamical simulations for modeling CE has allowed us to make an accurate determination of the total angular momentum content of the core, which we have robustly confirmed with analytical calculations (R. W. Everson et al. 2024). We uncover here that the NS ends up being surrounded by dense core material with significant rotational support to power a long-lived accretion transient, with features that are evocative of GRB progenitors.

In a pioneering study, R. A. Chevalier (2012), motivated by the calculations of C. L. Fryer & S. E. Woosley (1998) in relation to the merger between a BH and a massive helium star, postulated that CE events involving BHs or NSs could power strong explosions. The ensuing explosion, envisioned to be powered by mass accretion onto the compact object, is expected to easily unbind the entire envelope. In this scenario, the interaction of the rapidly moving envelope with the mass lost during the initial CE phase is predicted to produce a type IIn supernova. This scenario was thoughtfully extended by S. L. Schrøder et al. (2020). They used radiative transfer models of blast waves interacting with the CE ejecta in order to calculate the accompanying light curves. In both of these studies, the explosion driven by the coalescence of the compact object with the companion's core was parameterized using straightforward energetic arguments and assumed to be shared roughly spherically with the envelope.

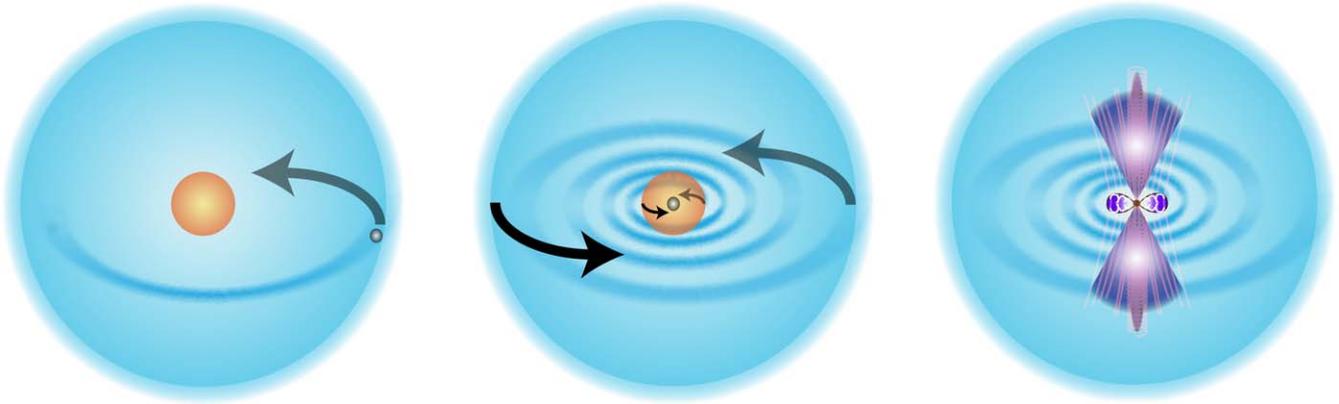

**Figure 11.** Diagram illustrating the post-merger evolution of the LMC X-4 system. After the primary fills its Roche lobe (represented by the blue sphere) and engulfs the NS (represented by the gray sphere), a merger event cannot be avoided (Figure 2). As the engulfment proceeds (Figure 7), the NS rapidly ($\approx 13.8$ hr) sinks into the core (represented by the orange sphere). As a result, the core spins up due to shocks produced in the course of the inspiral. The further evolution of the rapidly rotating merger product can produce several interesting phenomena. The neutron star is expected to accrete a significant amount of mass to collapse to a BH. The rotating core, as it continues to collapse, will produce an accretion disk (Figure 8), which will likely also power and launch a relativistic jet.

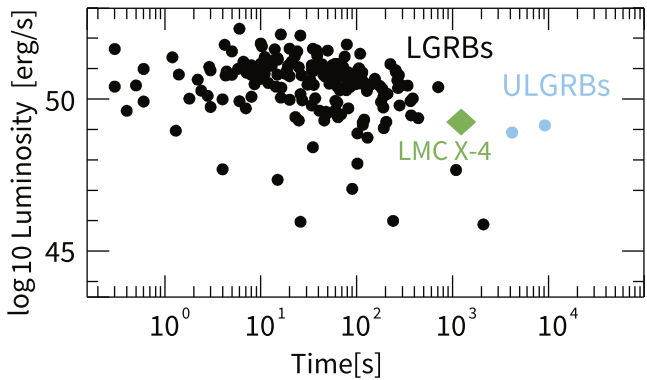

**Figure 12.** Luminosity as a function of the duration of high-energy transients adapted from A. J. Levan et al. (2014). For the GRB and ULGRB sources, we plot $t_{90}$ as the duration and the peak isotropic luminosity. The duration of the event assumes the ejection timescale determined by accretion feedback in Figure 9. The properties of the high-energy transient roughly coincide with those of the emerging class of ultra-long GRBs, such as GRB 101225A, GRB 111209A, and GRB 121027A.

Another relevant study explored the fate of TŻOs with massive envelopes (T. J. Moriya 2018), which are expected to be supported by nuclear reactions (R. C. Cannon 1993). T. J. Moriya (2018) examined the collapse of a TŻO after the star effectively runs out of fuel, as originally envisioned by P. Podsiadlowski et al. (1995). T. J. Moriya (2018) calculated the mass accretion onto the NS and concluded that a collapse to a BH was likely to occur. The consequences of this scenario depend sensitively on the angular momentum content of the envelope, which is assumed to be quasi-spherical (P. Podsiadlowski et al. 1995) and only likely to form a disk days after collapse (T. J. Moriya 2018). Even in the most optimistic conditions (i.e., assuming prompt disk formation), the collapse of a hydrostatic TŻO is expected to power a transient with corresponding mass accretion rates $\lesssim 10^{-7} M_\odot\, \mathrm{s}^{-1}$. As such, the luminosity of these transients is expected to be at least 5 orders of magnitude dimmer than those discussed in Section 4.2 (Figure 9). Yet, they should be a natural end state in the evolution of quasi-static TŻOs.

### 4.4. Thin-envelope TŻOs

After nearly all of the stellar envelope has been unbound and ejected by the nascent BH, for systems with compact primaries such as LMC X-4, an alternative type of steady-state merger product may be possible: the thin-envelope TŻO (TETŻO) as proposed by R. W. Everson et al. (2024). The TETŻO comprises a central BH accreting via a disk surrounded by a diffuse, radiation supported, remnant envelope with $\lesssim 1\%$ of the envelope's initial mass, which is powered by the BH's accretion luminosity. The TETŻO is not unlike the classical TŻO; although, rather than presenting similarly to a star in the optical bands, these would present as ultra-luminous X-ray sources with a lifetime of $\approx 10^4$ yr (R. W. Everson et al. 2024).

Thus, the fate of LMC X-4 is unlikely to lead to a configuration as a TŻO due to the effects of rotation as described above, but if any envelope remains after the series of transient events initiated by the merger, this and other close HMXBs may end their lives as something not altogether different from what Thorne & Żytkow envisioned nearly 50 yr ago.

## 5. Summary

In this paper we investigate the long-term fate of high-mass X-ray binaries with the objective of understanding whether a TŻO might be ultimately assembled. We present the results from a 3D hydrodynamical simulation aimed at studying the fate of LMC X-4, a tight high-mass X-ray binary system, after the primary star fills its Roche lobe and engulfs the NS companion (Figure 11). The salient findings of this work are as follows:

1. The NS, as a consequence of the merger, injects energy, mass, and angular momentum into the core of the primary star (Figure 3). Spiral shocks during the merger spin up the core of the primary star before the NS eventually settles into the core's interior (Figure 7). LMC X-4 thus plainly develops into a hyper-accreting NS surrounded by a rotating stellar core, similar to a collapsar-like progenitor system (Figure 8).

2. The inspiraling NS, upon merging with the core, can accrete efficiently at high rates (Figure 9), subsequently collapsing into a BH. This accretion process is expected to power and launch a relativistic jet, which will pierce through the envelope of the primary star and trigger a bright transient with a luminosity and duration typical of an ultra-long GRB (Figure 12). Such high-energy transients can be uncovered by observers along the jet axis.

3. Significant post-merger accretion feedback will unavoidably unbind the vast majority of the surrounding envelope, powering an optical or infrared transient as the material expands, recombines, and becomes transparent. The detection of this transient, which will be available to observers at all directions, would offer direct constraints on the conditions and flow properties of this highly uncertain phase of high-mass X-ray binary evolution.

4. After most of the stellar envelope is ejected by the newly formed BH, we expect a thin-envelope TŻO to form, as predicted by R. W. Everson et al. (2024). This short-lived source ($\approx 10^4$ yr) will manifest as an ultra-luminous X-ray source. As such, the canonical framework for TŻO formation via common envelope evolution needs to be revisited. This is due to the fact that the post-merger angular momentum content of the material is, for all merging binaries (R. W. Everson et al. 2024), high enough to break spherical symmetry and produce a centrifugally supported accretion disk.

## Acknowledgments

We would like to thank the reviewer for providing comments that improved our manuscript. We gratefully acknowledge S. Wu, A. Mannings, A. Hermosillo Ruiz, M. MacLeod, R. Foley, D. Coulter, and S. Schrøder for helpful discussions. T.H.-S. acknowledges the support of the University of California—Historically Black College and University (UC-HBCU) Fellowship. R.W.E. acknowledges the support of the University of California President's Dissertation-Year and Eugene V. Cota-Robles Fellowships, the Heising-Simons Foundation, the ARCS Foundation, and the Vera Rubin Presidential Chair at UCSC. This material is based upon work supported by the National Science Foundation Graduate Research Fellowship Program under grant No. 1339067. R.Y. is grateful for support from a Doctoral Fellowship from the University of California Institute for Mexico and the United States (UCMEXUS), a Texas Advanced Computing Center (TACC) Frontera Computational Science Fellowship, and a NASA FINESST award (21-ASTRO21-0068). E.R.-R. acknowledges support from the Heising-Simons Foundation and the National Science Foundation (2150255 and 2307710). Any opinions, findings, and conclusions or recommendations expressed in this material are those of the authors and do not necessarily reflect the views of the NSF. The 3D hydrodynamics software used in this work was developed in part by the DOE NNSA- and DOE Office of Science-supported Flash Center for Computational Science at the University of Chicago and the University of Rochester.

*Software:* FLASH (B. Fryxell et al. 2000), Python, MESA (B. Paxton et al. 2011, 2013, 2015, 2018, 2019), matplotlib (J. D. Hunter 2007), yt (M. J. Turk et al. 2011), NumPy (S. van der Walt et al. 2011), py_mesa_reader (B. Wolf & J. Schwab 2017).

## Appendix A
## Supplementary Details on the Initial Stellar Model and Global Simulations

In Figure 2, we plot the evolution of the stellar surface for the $18M_\odot$ stellar companion in LMC X-4, as calculated by MESA using the mean LMC metallicity of [Fe/H] = −0.42, from "today" to "CE onset." In Figure 13 we show the convective velocity according to mixing length theory, the specific nuclear energy generation rate, and the H and He mass-fraction profiles as a function of the radial coordinate. The MESA model corresponding to the primary's structure today (left panel) has a convective core mass of $7.9M_\odot$ and a radius of $1.5R_\odot$, consistent with M. Falanga et al.'s (2015) $1\sigma$ observational constraints. At the time of CE onset (right panel), the MESA model's convective core had a mass of $6.2M_\odot$ and radius of $1.3R_\odot$. The radial coordinate at which the $\alpha = (E_{\mathrm{grav}}/\Delta E_{\mathrm{orb}}) = 1$ condition is satisfied is shown as a vertical dashed line in the right panel of Figure 13.

When the system enters into a CE phase, the stellar companion has a luminosity of $6.8 \times 10^4 L_\odot$. Shortly after, the NS is found to be completely engulfed within the companion's nonconvective envelope. Owing to gas drag, the NS will spiral toward the companion's core. We follow the global hydrodynamic evolution from CE onset up to the NS settling into the stellar core in our FLASH simulation. Figures 3 and 14 show the NS inspiral phase in both the orbital and meridional planes, respectively. R. A. Chevalier (1993, 1996) argued that steady-state TŻOs may not be able to form since, during the inspiral phase, accretion onto the NS is likely to occur in the neutrino-dominated regime rather than the widely discussed Eddington-limited regime. In Appendix B that follows, we explore this assertion and investigate the applicability of the known hydrostatic solutions.

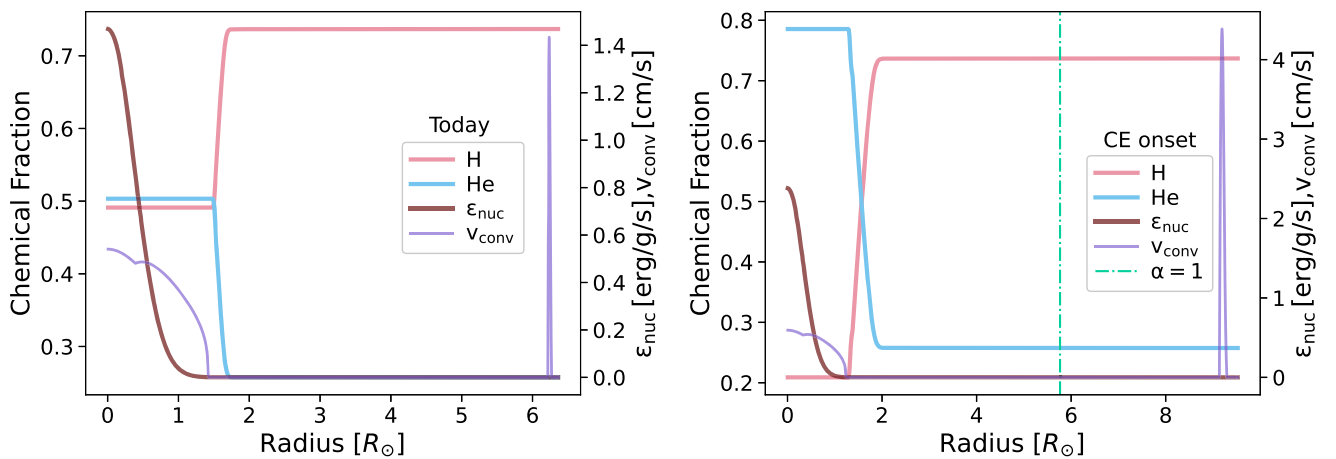

**Figure 13.** Stellar structure profiles for the $18M_\odot$ MESA model presented in Figure 2. Plotted are the specific nuclear energy generation rate (in units of $10^5$ erg g$^{-1}$ s$^{-1}$), the convective velocity according to mixing length theory (in units of $10^5$ cm s$^{-1}$), and the H and He mass-fraction profiles as a function of radial coordinate. The left and right panels show the interior properties of the companion as inferred "today" by M. Falanga et al. (2015) and at the time of mapping into the FLASH simulation, respectively. The radial coordinate at which the $\alpha = 1$ condition is satisfied is shown as a green vertical dashed line in the left panel.

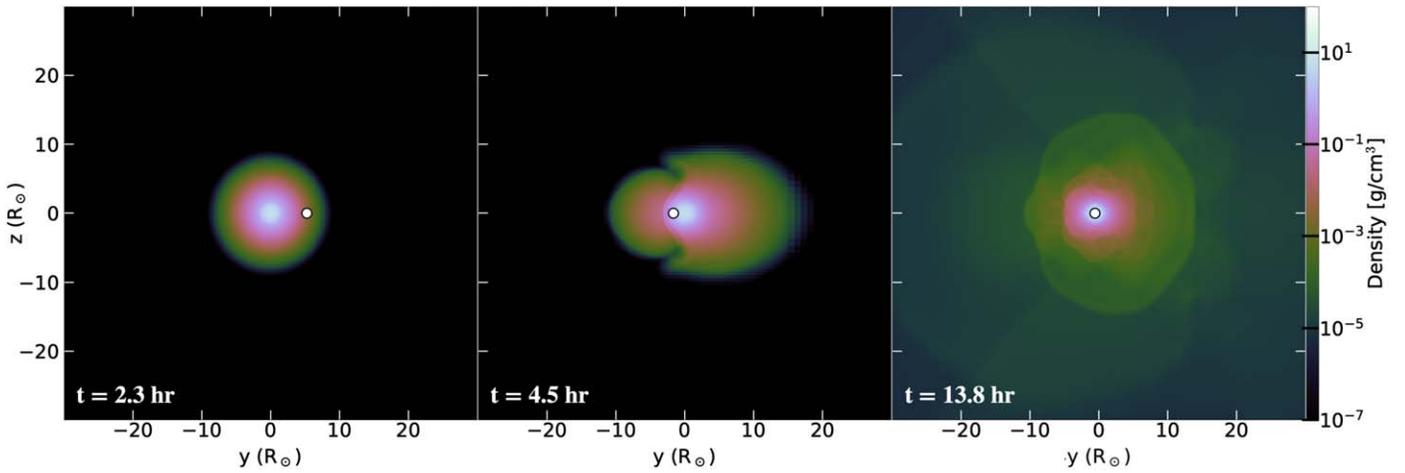

**Figure 14.** Two-dimensional slices of the logarithm of gas density in the meridional plane (y–z) at three different times during the inspiral phase (2.3 hr, 4.5 hr, and 13.8 hr). The upper panels in Figure 3 show the corresponding density slices in the orbital plane (y–x).

## Appendix B
## On the Applicability of Equilibrium TŻO Solutions

HMXB LMC X-4 consists of an NS orbiting an O8 III companion star with $L \approx 5 \times 10^4 L_\odot$ in a 1.4 days orbit (M. Falanga et al. 2015). LMC X-4 has a characteristic X-ray luminosity of $L_x \approx 2 \times 10^{38}$ erg s$^{-1} \approx 10^5 L_\odot$, close to the Eddington limit for neutron stars. When the massive stellar companion completely fills its Roche lobe (Figure 2), the mass-transfer rate onto the neutron star will exceed the Eddington limit ($\approx 10^{-8} M_\odot$ yr$^{-1}$) by several orders of magnitude. Unless the majority of the energy goes into neutrinos, the neutron star will not be able to accrete most of the transferred mass, such that the excess mass will form a quasi-hydrostatic structure around the NS (R. A. Chevalier 1993).

There are two main difficulties to reach the classical equilibrium solutions initially envisioned by K. S. Thorne & A. N. Żytkow (1977) and R. C. Cannon (1993). First, the motion of the NS through the envelope will sweep away any equilibrium structure that might form around the NS. Second, as the NS sinks into the stellar core, the mass accretion is expected to increase much faster than the time it takes hydrostatic equilibrium to be attained. Not to mention that the mass accretion rate will unavoidably exceed $10^1 M_\odot$ yr$^{-1}$ (Figure 6) and, under these circumstances, rapid accretion with significant neutrino losses cannot be prevented (R. A. Chevalier 1996; C. L. Fryer et al. 1996). Be that as it may, we note as a caveat that a self-consistent calculation of the formation of a TŻO has not been performed, in particular, at accretion rates between the Eddington limit, $\dot{M}_{\rm Edd} \approx 10^{-8} M_\odot$ yr$^{-1}$, and the neutrino-cooling limit $\dot{M}_\nu \approx 10^4 \dot{M}_{\rm Edd} \approx 10^{-4} M_\odot$ yr$^{-1}$ (J. C. Houck & R. A. Chevalier 1991). In what follows, we show that the formation of TŻOs during CE considerably bridges this luminosity gap.

Two types of hydrostatic TŻO solutions arise based on the type of energy source at the base of the compact core. For TŻOs embedded in moderately massive envelopes, the main energy source is thought to be limited by the self-regulatory balance between gravity and radiation pressure (R. C. Cannon 1993). In more massive envelopes, the generation of $e^- - e^+$ pairs results in a reduction of the Eddington limit, and the luminosity necessary to satisfy stability must therefore be made up by nuclear reactions (G. T. Biehle 1991, 1994; R. C. Cannon 1993). There is a gap in luminosity between the two types of solutions.[13] Envelopes supported by accretion have luminosities between $10^4$ and $5 \times 10^4 L_\odot$ while envelopes supported by nuclear reactions attain luminosities very close to $10^5 L_\odot$.

Figure 15 shows how temperature varies with density at the base of a given steady-state TŻO solution as calculated by R. C. Cannon (1993). For these cases, the luminosity and temperature are connected simply through the reaction rates. The steady-burning phase in this regime is expected to be terminated when the supply of seed elements is exhausted and the rp-process becomes inefficient. This depends strongly on convection, which is rather uncertain under these conditions. It is, however, expected that as the accretion rate increases by orders of magnitude above the Eddington rate, a radiative zone is likely to develop at the base of the envelope, and the supply of fuel to the burning region is expected to be cut off (R. A. Chevalier 1996; C. L. Fryer et al. 1996).

As the mass accretion rate continues to increase, the core region progressively heats up until neutrino losses become the dominant energy mechanism, and neutrino runaway becomes inescapable (J. C. Houck & R. A. Chevalier 1991). In Figure 15 we show the thermodynamic properties along the inspiral trajectory of the NS, as inferred using our global hydro-dynamical calculations. The $\rho - T$ conditions at the point-mass accretion boundary ($\approx 0.1 R_\odot$), which is significantly smaller than the gravitational radius, are extrapolated to the NS surface using the analytical solutions derived by R. A. Chevalier (1996). The NS inspiral trajectory begins, as expected, at significantly higher densities than the ones supported by the steady-state TŻO solutions. The temperature and density at the start of the inspiral phase are near the limit in which pair emissivity dominates the neutrino emission.[14] As soon as the accretion rate reaches $\gtrsim 10^{-1} M_\odot \, \mathrm{yr}^{-1}$, neutrino runaway becomes inescapable. This is observed to take place at around 2.3 hr (Figures 3 and 14). In brief, the considerations given here show that even during the inspiral phase, the TŻO equilibrium conditions are difficult to reach. Once the NS settles into the core and collapse ensues, the corresponding accretion rates are reminiscent of those expected to take place in gamma-ray burst progenitor systems (C. L. Fryer et al. 1999; R. Popham et al. 1999).

---

[13] We note that R. Farmer et al. (2023) neglected energy generation by nuclear heating in their static TŻO models.

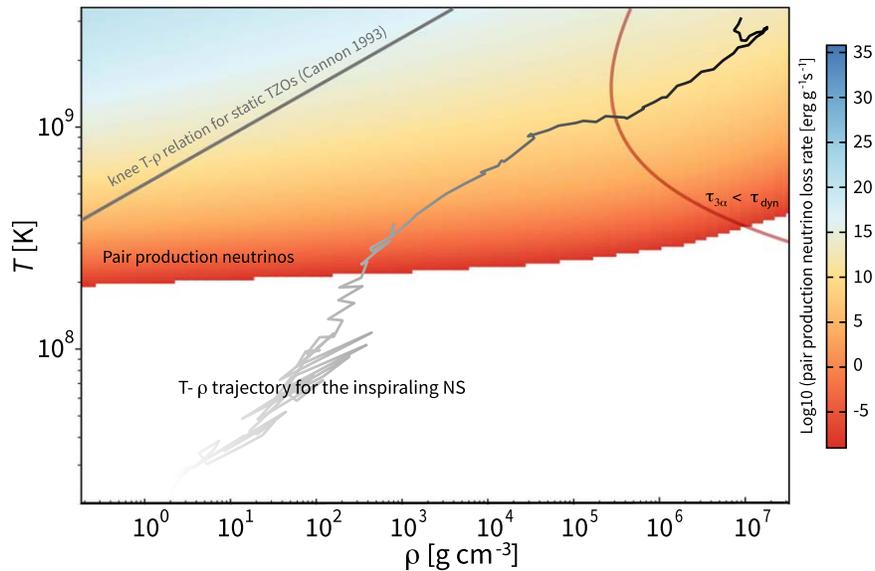

**Figure 15.** $\rho T$ parameter space relevant to TŻO hydrostatic solutions and the inspiral phase of an NS during CE. The gray line shows how temperature varies with density at the base of static TŻO solutions (R. C. Cannon 1993). The inspiral trajectory starts at comparatively low temperature and less dense material (lower corner) and becomes hotter and denser (upper corner) as the NS sinks into the core. The neutrino emission is dominated by electron-positron pair production processes, which are computed for solar metallicity material according to the fit formulae of N. Itoh et al. (1996).

Figure 15 also presents the $\rho - T$ parameter space relevant to He ignition. This is particularly relevant when the NS enters the companion's core region. In this region, the timescale on which the He material can react, $\tau_{\mathrm{dyn}} = (G\rho)^{-1/2}$, starts being comparable to the burning timescale (A. M. Khokhlov & E. V. Ergma 1986), and the material cannot expand swiftly enough to quench burning (S. Rosswog et al. 2009; C. Holcomb et al. 2013). The fate of the envelope material is mainly governed by the competition between dynamical nuclear heating and neutrino cooling (S. Rosswog et al. 2008). This regime was initially explored by C. Fryer et al. (1999) and more recently by R. Fernández & B. D. Metzger (2013). The broad implication is that the energy associated with dynamical burning is not expected to prevent the NS merging with the companion's core or to significantly alter the accretion flow during the collapse of the companion's core onto the NS (A. I. MacFadyen et al. 2001).

As we have described here, our rationalization of the principal physical considerations of the formation of TŻOs via CE evolution combines some generally accepted features with some more speculative physical ingredients. What is more valuable, though considerably harder to achieve, is to refine models like the one presented here to the point of resolving the properties of the flow near the accreting NS. What we can hope of our current modeling is that it will help us in this endeavor.

### ORCID iDs

Tenley Hutchinson-Smith ● https://orcid.org/0000-0002-3472-2453
Rosa Wallace Everson ● https://orcid.org/0000-0001-5256-3620
Angela A. Twum ● https://orcid.org/0009-0006-4675-7596
Aldo Batta ● https://orcid.org/0000-0002-3269-3847
Ricardo Yarza ● https://orcid.org/0000-0003-0381-1039
Jamie A. P. Law-Smith ● https://orcid.org/0000-0001-8825-4790
Alejandro Vigna-Gómez ● https://orcid.org/0000-0003-1817-3586
Enrico Ramirez-Ruiz ● https://orcid.org/0000-0003-2558-3102

---

[14] The neutrino emission is computed for solar metallicity material according to the fit formulae of N. Itoh et al. (1996) as coded by F. Timmes: https://cococubed.com/.